\documentstyle[12pt]{article}
\date{}
\def\be{\begin{equation}}
\def\ee{\end{equation}}
\def\bea{\begin{eqnarray}}
\def\eea{\end{eqnarray}}
\def\L{\Lambda}
\def\sn{\sqrt n}

\title{Multidimensional cosmological solutions
of Friedmann type in dilaton gravity}
\author{G.\,S. Sharov\thanks{E-mail: german.sharov@tversu.ru},
E.\,G. Vorontsova\\
{\small Tver state university}\\
{\small Tver, 170002, Sadovyj per. 35, Mathem. dep-t.}}
\begin{document}
\maketitle
\begin{abstract}
In $D$\,-\,dimensional dilaton gravitational model
with the central charge deficit $\L$ the generalized Friedmann-type
cosmological solutions (spatially homogeneous and isotropic)
are obtained and classified.
\end{abstract}
\bigskip
\noindent{\large\bf Introduction}
\medskip

One of the consequences of the string theory in its low-energy limit
is the scalar dilaton field $\phi(x)$ \cite{FradTseyt,GSW} in the space-time
with the (critical) dimension $D$. The dilaton was involved
in various gravitational models. In particular, it makes non-trivial
the $1+1$\,-\,dimensional (2D) gravity that has fruitfully progressed
during the last ten years \cite{CallanGHS,GibbonsP,PelzerS}.

On the other hand the dilaton field and other lowest-order
corrections to Einstein gravity from the string theory
were intensively used in cosmology \cite{BailinL}\,--\,\cite{BarreirCC}.
Some of the mentioned authors \cite{CasasGQ}\,--\,\cite{BarreirCC}
searched $3+1$\,-\,dimensional solutions
with the Friedmann-Robertson-Walker metric keeping in mind
some form of compactification of other $D-4$ dimensions.
The dilaton field was recognized as a mechanism for the
inflation \cite{infl} that was required for solving some
cosmological problems.
In Refs.~\cite{GasperVenez93A,BruVenez,EasthM96,KaloperMO} the dilaton
was used in attempts to avoid the Big Bang singularity.

A. Tseytlin \cite{Tseyt92} obtained a set of cosmological solutions
for the maximally symmetric space of an arbitrary dimension $D$
with the dilaton, the central charge and the covariantly constant antisymmetric
tensor field (for $D=3$ or 4).

The latter approach is used in the present paper.
We consider and classify cosmological solutions in
$D$\,-\,dimensional dilaton gravity with the central charge deficit $\L$
for the case of an arbitrary $D$. We use the $D$\,-\,dimensional
generalization of FRW metric corresponding to homogeneous and
isotropic spatial part with the constant curvature ($k=0,\,\pm1$).
All spatial dimensions are equal as it may be before the compactification.
The purpose of our studying is to order the known $2D$
and higher $D$ dilaton cosmological solutions
(with adding new ones) and to classify them exhaustively.

In this paper we  don't consider the terms in the
action with the antisymmetric tensor fields
so the condition of the maximal spatial symmetry
restricts the possible dimensions within $D=3$
and $D=4$. In these cases the problem was studied by many authors
\cite{BailinL,Tseyt92,BehrndtF94,CopelLW94,EasthMW96,CopelEW}.

\bigskip
\noindent{\large\bf 1. Model and equations}
\medskip

The dilaton gravitational model with the following action \cite{FradTseyt}
is considered:
\be
S=\int\sqrt{|g|}\,e^{-2\phi(x)}\left[R+4(\nabla\phi)^2+\L
\right]\,d^{n+1}x.
\label{S}\ee
Here $\phi(x)$ is the scalar dilaton field in $n+1$\,-\,dimensional
($D=n+1$) pseudoriemann manifold $M_{1,n}$ with coordinates $x^\mu$
and the metric tensor $g_{\mu\nu}$, $g={}$det\,$|g_{\mu\nu}|$, $R$ is the
scalar curvature, $(\nabla\phi)^2=g^{\mu\nu}\partial_\mu\phi\partial_\nu\phi$.

We use here and below the string frame for the action (\ref{S})
keeping in mind that it may be conformally transformed
$g_{\mu\nu}\;\to\;e^{-r\phi}g_{\mu\nu}$, $r=4/(D-2)$ to the Einstein frame
\cite{FradTseyt,Tseyt92,GasperVenez93}.

Equations of evolution for this model
$$\begin{array}{c}
R_{\mu\nu}-\frac12(R+\L)\,g_{\mu\nu}+2\phi_{;\mu\nu}=
g_{\mu\nu}\left[2\phi_{;\lambda}^{;\lambda}-2(\nabla\phi)^2\right],\\
\frac14(R+\L)+\phi_{;\lambda}^{;\lambda}-(\nabla\phi)^2=0
\rule[3mm]{0mm}{1mm}
\end{array}$$
are obtained from action (\ref{S}) by the variation
of the gravitational field $g_{\mu\nu}$ and the dilaton field $\phi$
accordingly. Here $R_{\mu\nu}$ is the Ricci tensor, $\phi_{;\mu\nu}$ is the
covariant derivative.

Transform the latter equations to the equivalent system
\be\begin{array}{c}
R_{\mu\nu}+2\phi_{;\mu\nu}=0, \\
R+4(\nabla\phi)^2=\L.\rule[3mm]{0mm}{1mm}\end{array}
\label{sys}\ee

We search cosmological Friedmann's type solutions of Eqs.~(\ref{sys}).
It means that $M_{1,n}=R\times S^n_k$ where $S^n_k$ is the homogeneous and
isotropic manifold described by the scale factor $a(t)$ \cite{Tseyt92}.
The manifold $S^n_k$ has one of the following forms:
$n$\,-\,dimensional sphere $S^n_1$, pseudosphere $S^n_{-1}$, flat space
$S^n_0\equiv R^n$. In the spherical ($k=1$) case $a$ is the radius
of $S^n_1$. The metric in $M_{1,n}=R\times S^n_k$ may be taken
in the form \cite{Sh}
\be
ds^2=dt^2-a^2(t)\bigl[(dx^1)^2+c^2_k(x^1)(dx^2)^2+\dots
+c^2_k(x^1)\cdot c^2_k(x^2)\dots\cdot c^2_k(x^{n-1})(dx^n)^2\bigr],
\label{metr}\ee
where \,$c_k(x)=\left\{\begin{array}{ll}\cos x,& k=1, \\
1, &k=0, \\
\cosh x,&k=-1.\end{array}\right. $

We substitute the metric (\ref{metr}) and $\phi=\phi(t)$ into Eqs.~(\ref{sys})
taking into account the following expressions (for all $k$):
\bea
&R^0_0={-na\ddot a}/{a^2},\quad R^1_1=R^2_2=\dots=R^n_n
=-\left [{\ddot a}{a}+(n-1)({\dot a}^2+k)\right]/{a^2},& \nonumber\\
&R={-n\left[2a\ddot a+(n-1)({\dot a}^2+k)\right]}/{a^2},\quad\dot a=da/dt.&
\nonumber\eea
It reduces system (\ref{sys}) to three nontrivial equations
\bea
&-n\ddot a/a + 2\ddot\phi=0,& \label{eq1}\\
&-\ddot a-(n-1)({\dot a}^2+k)/a+2\dot a\dot\phi=0,& \label{eq2}\\
&n(n-1)({\dot a}^2+k)/a^2+4{\dot\phi}^2-4n\dot\phi\dot a/a=\L.& \label{eq3}
\eea

These three equations are not independent. Eq.~(\ref{eq3})
is the  ``zero energy" constraint that is conserved as the consequence of
Eqs.~(\ref{eq1}) and (\ref{eq2}) and hence gives only a restriction on
the initial values of $a$, $\dot a$, $\phi$, $\dot\phi$ \cite{Tseyt92}.

One may express $\dot\phi$ from Eq.~(\ref{eq2})
\be
\dot\phi=\frac12\left[\frac{\ddot a}{\dot a}+\frac{n-1}{a\dot a}
({\dot a}^2+k)\right]
\label{phi}\ee
and substitute it into Eq.~(\ref{eq3}). The resulting second-order equation
(it is quadratic one with respect to $\ddot a$) may be transformed through
the substitution
\be \dot a=q(a)
\label{aq}\ee
to the following form:
\be
\frac{dq}{da}=\frac qa-k\frac{n-1}{aq}\pm\sqrt{n\frac{q^2}{a^2}
-kn\frac{n-1}{a^2}+\L}.
\label{qgen}\ee
To classify all solutions of the latter equation and system
(\ref{eq1})\,--\,(\ref{eq3}) we consider some cases separately.

\bigskip
\noindent{\large\bf 2. Flat universe: $k=0$}
\medskip

In this case Eq.~(\ref{qgen}) is homogeneous one. It may be integrated
for all values of $\L$:
$$
q=\frac a2\frac{C^2a^{\pm2\sn}-\L/n}{Ca^{\pm\sn}}.
$$
It results in the following evolution of the scale factor:
\be
a(t)=a_0\cdot\left\{\begin{array}{ll}
\Big[\tanh(\frac12\sqrt{\L}\,\tau)\Big]^{\pm1/\sn}, & \L>0, \\
\,\tau^{\pm1/\sn},& \L=0,\\
\Big[\tan(\frac12\sqrt{-\L}\,\tau)\Big]^{\pm1/\sn}, &\L<0.\end{array}\right.
\label{ak0}\ee
Here $C$ or its equivalent $a_0$ ($a_0^{\mp2\sn}=nC^2/|\L|$ for $\L\ne0$)
is the constant of integration in (\ref{qgen}). We use in Eq.~(\ref{ak0})
and below the time parameter
\be
\tau=\pm(t-t_0)=|t-t_0|,
\label{tau}\ee
because all solutions of system (\ref{eq1})\,--\,(\ref{eq3}) are invariant
with respect to time translation
($t_0$ is the constant of integration) and time reflection.
The signs $\pm$ in Eqs.~(\ref{tau}) and (\ref{ak0}) (the latter is connected
with the sign in Eq.~(\ref{qgen})) are independent.

The expression for $\phi$ corresponding to (\ref{ak0}) results from
Eq.~(\ref{phi})
\be
e^{2\phi-2\phi_0}=\left\{\begin{array}{ll}
\Big[\sinh(\frac12\sqrt{\L}\,\tau)\Big]^{-1\pm\sn}
\Big[\cosh(\frac12\sqrt{\L}\,\tau)\Big]^{-1\mp\sn}, & \L>0, \\
\;\tau^{-1\pm\sn},& \L=0,\\
\Big[\sin(\frac12\sqrt{-\L}\,\tau)\Big]^{-1\pm\sn}
\Big[\cos(\frac12\sqrt{-\L}\,\tau)\Big]^{-1\mp\sn}, &\L<0.\end{array}\right.
\label{phik0}\ee

The functions $a(t)$ (\ref{ak0}) and $\phi(t)$ (\ref{phik0}) are exact
solutions not only of Eqs.~(\ref{eq2}) and (\ref{eq3}) but also of
Eq.~(\ref{eq1}). This situation is more general.
It is the non-trivial fact connected with the constraint character of
Eq.~(\ref{eq3}) that all solutions of Eqs.~(\ref{qgen}) for all $k$, $\L$
and $n$ satisfy Eq.~(\ref{eq1}).

It is necessary to supplement the family (\ref{ak0}),
(\ref{phik0}) by the linear dilaton solution \cite{MyersAntonBEN,KaloperMO}
\be
a={}\mbox{const},\quad \phi=\phi_0\pm\frac12\sqrt{\L}\,t,\qquad \L\ge0,
\quad k=0,\label{lin}\ee
that was lost because of dividing on $\dot a$ in Eq.~(\ref{phi}).

The solutions (\ref{ak0}), (\ref{phik0}) are well known in the $4D$ or
$n=3$ case \cite{BruVenez,EasthMW96,KaloperMO} and were considered for $\L>0$
and an arbitrary $D$ in Ref.~\cite{Tseyt92}. They coincide with the solutions
obtained by Mueller \cite{Mueller} for the manifold $R\times T^n$ in the case
when all radii of the n-dimensional torus $T^n$ are equal. The solutions
of Veneziano \cite{Venez91} for the flat manifold with different scale factors
generalize Eqs.~(\ref{ak0}), (\ref{phik0}).

One can see that qualitative picture of solutions (\ref{ak0}), (\ref{phik0})
and the structure of branches
are similar for all $n>1$. In the cases $\L\ge0$ for all $n$ we have
two different branches of solutions (or 4 branches if we take two symmetric
with respect to the change ``$+$" to ``$-$" in (\ref{tau}) solutions
as different ones \cite{BruVenez})\footnote{Below for counting a number of
the branches we don't differ two solutions connected only by the time
reversal (\ref{tau}). Similarly, describing an expansion we always keep
in mind that the corresponding solution with the contraction exists.
If one differ these solutions he should multiply any mentioned number
of branches by two.}.
And there are only one branch for $\L<0$ due to the symmetry properties
of (\ref{tau}).

The upper sign in Eqs.~(\ref{ak0}) and (\ref{phik0}) corresponds to
an expansion (with increasing $\tau$) of the flat dilatonic universe from
the singular point $\tau=0$ with $\phi$ growing up from $-\infty$,
the opposite sign describes an evolution without
the singularity $a=0$ but with $a\to\infty$, $\phi\to+\infty$ at $\tau\to0$.
The value $\L=0$ always corresponds to the power-law expressions for $a$
and $\phi$ generalizing ones \cite{BruVenez} for the case $n=3$.
The presence of $\L>0$ results in slowing down the above mentioned
expansion and stabilization of the scale factor:
$a\to a_0={}$const, $\phi\to-\infty$ at $\tau\to\infty$.
In the case $\L<0$ we, v.v. have acceleration of evolution that is
finished during finite time.

Thus the described above role of $\L$-term in the dilaton cosmology
is in some sense opposite \cite{Tseyt92} to that in usual Friedmann type
$n+1$\,-\,dimensional cosmology. In the latter case the evolution of
the flat universe ($k=0$) with dust matter\footnote{In action (\ref{S})
all matter is in ``exotic" (dilaton and $\L$-term) form.} and with Einstein's
gravitational Lagrangian $\sqrt{|g|}\,(R+2\L)$ has the form \cite{Sh}
$$
a={\mbox{const}}\cdot\left\{\begin{array}{ll}
\Big[\sinh\big(\sqrt{\frac{n\L}{2(n-1)}}\,\tau\big)\Big]^{2/n}, & \L>0, \\
\,\tau^{2/n},& \L=0,\\
\Big[\sin\big(\sqrt{\frac{-n\L}{2(n-1)}}\,\tau\big)\Big]^{2/n}, &\L<0,
\end{array}\right.
$$
and $\L>0$ accelerates the expansion.

Note that for the $2D$ dilaton model ($n=1$) all ``cosmological" solutions
are exhausted by Eqs.~(\ref{ak0})\,--\,(\ref{lin}). It is naturally so the
$1$\,-\,dimensional space has no an intrinsic curvature, hence $k$ vanishes
in Eqs.~(\ref{eq2}), (\ref{eq3}) if $n=1$. In this case the $n=1$ solution
(\ref{ak0}), (\ref{phik0}) coincides with the $D=2$ string ``black hole"
solution \cite{GibbonsP,TseytV92,Witten1} (taking the Euclidean form after
the analytic continuation $\tau\to ir$, $\L\to-\L$ \cite{Tseyt91}).

\bigskip
\noindent{\large\bf 3. Curved space $k=\pm1$ and $\L=0$}
\medskip

For any value of $k$ all solutions of system (\ref{eq1})\,--\,(\ref{eq3})
may be obtained by integrating Eq.~(\ref{qgen}). It was mentioned above
that Eq.~(\ref{eq1}) in this case is satisfied too. But the most
convenient way to classify various types of the solutions for $\L=0$
is to exclude the scale factor $a$ from Eq.~(\ref{eq1})\,--\,(\ref{eq3}).
For this purpose we take the linear combination of these equations
with the coefficients $-1$, $n/a$ and 1. The resulting expression
\be
\frac{\dot a}a=\frac1n\Big(2\dot\phi-\frac{\ddot\phi}{\dot\phi}\Big)
\label{h}\ee
(with $\L=0$) is integrable:
\be
a^n=\tilde Ce^{2\phi}/\dot\phi.
\label{a}\ee
Here $\tilde C$ is an arbitrary constant.

Before using the expressions (\ref{h}) or (\ref{a}) we are to consider
all solutions of system (\ref{eq1})\,--\,(\ref{eq3}) with $\dot\phi=0$.
For all $k$, $\L$, $n>1$ we have only two solutions of this type:
the trivial case of (\ref{lin}) $a=a_0$, $\phi=\phi_0$, $k=0$, $\L=0$
and the solution with linear grows of $a$ \cite{Tseyt92}
\be
a=a_0\pm t,\quad\phi=\phi_0,\qquad k=-1,\quad \L=0.
\label{alin}\ee

Substituting Eq.~(\ref{h}) or (\ref{a}) into Eq.~(\ref{eq1})
we obtain the differential equation
\be
n\frac d{dt}\frac{\ddot\phi}{\dot\phi}=
\Big(2\dot\phi-\frac{\ddot\phi}{\dot\phi}\Big)^2.
\label{eqphi}\ee
Introducing the notation
$$p=\ddot\phi/\dot\phi^2$$
we integrate Eq.~(\ref{eqphi}) in the following form:
\be
\frac{d\phi}{dt}=\frac1Q\left|p+\frac2{\sn-1}\right|^{\frac{-
\sn}{2(\sn-1)}}\left|p-\frac2{\sn+1}\right|^{\frac{-\sn}
{2(\sn+1)}}.
\label{dphi}\ee
Here $Q$ is the constant of integration that can take various
values in three intervals resulting from the division of the axis
$-\infty<p<\infty$ by the singular points
\be
p_1=-\frac2{\sn-1},\qquad p_2=\frac2{\sn+1}.
\label{p12}\ee

Taking into account that $p\,dt=-d\,\big(\dot\phi\big)^{-1}$
we obtain the relation
\be
dt=-\frac{Qn}{n-1}\left|p-p_1\right|^{\frac{2-\sn}{2(\sn-1)}}
\left|p-p_2\right|^{\frac{-2-\sn}{2(\sn+1)}}
\mbox{sign}\,(p-p_1)\,\mbox{sign}\,(p-p_2)\,dp.
\label{dt}\ee
It expresses $t$ through the parameter $p$ in quadratures
in each of the intervals $(-\infty,p_1)$, $(p_1,p_2)$, $(p_2,+\infty)$.

Eqs. (\ref{a}), (\ref{dphi}) and (\ref{dt}) let us express $\phi$ and
$a$ through $p$:
\bea
a(p)&=&C_a\big|p-p_1\big|^{\frac1{2(\sn-1)}}
\big|p-p_2\big|^{\frac{-1}{2(\sn+1)}},\label{ap}\\
\phi(p)&=&\phi_0+\frac{\sn}4\ln\left|\frac{p-p_1}{p-p_2}
\right|.\label{phip}
\eea
Here the constants $C_a$, $\phi_0$ and $\tilde C$ in Eq. (\ref{a})
are connected: $C_a=|\tilde CQe^{2\phi_0}|^{1/n}$.

Expressions (\ref{dt})\,-\,(\ref{phip}) satisfy
Eqs.~(\ref{eq1})\,--\,(\ref{eq3}) only under the following condition:
\be
(nQ/C_a)^2=-k\,\mbox{sign}\,(p-p_1)\cdot{\mbox{sign}}\,(p-p_2).
\label{signs}\ee
It means that the solution (\ref{dt})\,-\,(\ref{phip}) for the spherical
case of $k=1$ exist only if $p\in(p_1,p_2)$ and for the case $k=-1$ it
takes place if $p\in(-\infty,p_1)\cup(p_2,+\infty)$.

So for $k=1$, $\L=0$ there is only one branch of solutions (we count
the branches in accordance with the footnote in Sect.~2) represented
in Fig.~1 as the evolution of the scale factor $a$ in time (\ref{tau})
for various $n=D-1$ (solid lines). The graphs for $\phi=\phi(\tau)$
with $\phi_0=3$ are drawn by the dashed lines for the cases $n=2$ and $n=9$.
For any $n$ these solutions describe the
expanding universe ($a$ grows up from the singular value $0$
to $+\infty$ and $\phi$ rolls up from $-\infty$ to $+\infty$
when $p$ changes between the points (\ref{p12}) $p_1$ and $p_2$)
during the finite time
\be
T=\int\limits_{p_1}^{p_2}\frac{dt}{dp}\,dp=
Qn\Big(\frac{4\sn}{n-1}\Big)^{\frac1{n-1}}\Gamma\Big(\frac{\sn}{2(\sn-1)}\Big)
\Gamma\Big(\frac{\sn}{2(\sn+1)}\Big)\bigg/\Gamma\Big(\frac1{n-1}\Big).
\label{T}\ee
Here $\Gamma(x)$ is the Euler gamma function. In Fig.~1 the graphs $a=a(\tau)$
and $\phi=\phi(\tau)$ with the equal values $T=1$ for different $n$ are drawn.
Note that the solutions (\ref{dt})\,-\,(\ref{ap}) for any fixed $n>1$
and $k=\pm1$ constitute the two-parameter family: the first parameter $t_0$
in (\ref{tau}) is trivial (a translation in $t$), the second is $C_a$, or
$Q$, or $T$ (they are connected by Eqs.~(\ref{signs}) and (\ref{T})).
The third parameter $\phi_0$ contains Eq.~(\ref{phip}).

In the pseudospherical case $k=-1$ the solutions (\ref{dt})\,-\,(\ref{phip})
are divided into two branches which are shown in Fig.~2. The first one
(Fig.~2a) describes the expansion of $a$ and growing $\phi$:
$0<a<+\infty$, $-\infty<\phi<\phi_0$, $p<p_1$. The second branch
with $p>p_2$ (Fig.~2b) corresponds to evolution of $a$ without the
singularity $a=0$: the value $a$ decreases from $+\infty$ to the minimum
$a_{min}=C_a\big[|\sn\,p_1|^{p_1}(\sn\,p_2)^{p_2}\big]^{-1/4}$
and then it grows up to $+\infty$ when $\phi$ monotonically decreases
from $+\infty$ to $\phi_0$. In these both cases the asymptotic
$\tau\to\infty$ behavior is $\phi\to\phi_0={}$const,
$a\to{\mbox{const}}+\tau$, i.e. in this limit all solutions
(\ref{dt})\,-\,(\ref{ap}) for $k=1$ tend to the linear evolution
(\ref{alin}).

In the opposite (singular) limit $p\to p_1$ or $p\to p_2$ for $k=\pm1$
and for all branches we have an asymptotic dependence of the following two
types:
\bea
a\simeq{\mbox{const}}\cdot\tau^{1/\sn},&\quad&
e^{2\phi}\simeq{\mbox{const}}\cdot\tau^{-1+\sn},\quad\tau\to+0,
\label{lim0}\\
a\simeq{\mbox{const}}\cdot\tau^{-1/\sn},&\quad&
e^{2\phi}\simeq{\mbox{const}}\cdot\tau^{-1-\sn},\quad\tau\to+0.
\label{liminf}\eea
Eq.~(\ref{lim0}) corresponds to the case $p\to p_1$ and $a\to0$ that
takes place for the branches represented in Figs.~1 ($k=1$) and
2a ($k=-1$). The case (\ref{liminf}) is realized for the solutions in Fig.~2b
and in the limit $t\to T$ in Fig.~1.

We see that the asymptotic behavior of the solutions
(\ref{dt})\,-\,(\ref{phip}) ($k=\pm1$, $\L=0$) and
(\ref{ak0})\,-\,(\ref{phik0}) ($k=0$) is identical. So one can
easily set the correspondence between the equivalent branches
with the same limits (\ref{lim0}) or (\ref{liminf}).

There is only one\footnote{For the critical dimensions $D=10$ or
$D=26$ the exponents in (\ref{dt}) are rational but this expression
isn't integrable through any Euler's substitution.}
dimension $n=4$ (or $D=5$)
where an explicit solution (\ref{dt})\,-\,(\ref{phip}) exists:

\be
a(t)=\left|\frac{\tau^3-T^3}{\tau}\right|^{1/2},\quad
\phi(t)=C_{\phi}+\frac12\ln\left|\frac{\tau^3-T^3}{\tau^3}\right|.
\label{n4}\ee
Here $T=8Q/\sqrt[3]{3}$ in accordance with (\ref{T}),
$\tau=\pm(t-t_0)$ may be negative (unlike (\ref{tau})).

Expression (\ref{n4}) is rather simple and may illustrate the described
above solution (\ref{dt})\,-\,(\ref{phip}). The constant $t_0$ in
Eq.~(\ref{n4}) is chosen so that the point $\tau=0$ is the image of
$p=p_2=2/3$ and $\tau=T$ is the image of $p=p_1=-2$ for the continuous
mapping the $p$-axis onto $\tau$-axis. So the points $\tau=0$ and
$\tau=T$ divide the $\tau$-axis into 3 parts, and the interval
$0<\tau<T$ corresponds to the case $k=1$.

\bigskip
\noindent{\large\bf 4. Curved space $k=\pm1$ and $\L\ne0$}
\medskip

In these cases we study various types of evolution $a(t)$, $\phi(t)$
by solving the system (\ref{aq}), (\ref{qgen}) that may be rewritten as
\be
\dot a=q,\quad
\dot q=a^{-1}\left[q^2-k(n-1)\pm q\sqrt{n[q^2-k(n-1)]+\L a^2}\right].
\label{aqs}\ee

To consider solutions of this system we use in this Section, in particular,
numerical analysis so we have explicit solutions only in special cases.

For example, for the dimensionality $n=4$ or $D=5$ (specified in the previous
Section) and $\L>0$ we obtained the exact particular solutions with the
power-law evolution of the scale factor
\be
a=a_0\sqrt\tau,\quad e^{2\phi-2\phi_0}=\tau\cdot e^{k\sqrt\L\,\tau},\quad
a_0=\sqrt[4]{36/\L},\;\;n=4,\;\;k=\pm1,\;\;\L>0.
\label{asqrt}\ee

For other values $n$ the analogs of the solutions (\ref{asqrt}) also exist
but the expressions
\be
a\simeq a_0\sqrt\tau,\quad e^{2\phi-2\phi_0}\simeq
\tau^{(n-2)/2}\cdot e^{k\sqrt\L\,\tau},\quad
a_0^2=2(n-1)/\sqrt\L,\;\;k=\pm1,\;\;\L>0.
\label{asqn}\ee
are only asymptotic at $\tau\to\infty$ \cite{Tseyt92}.
In the opposite limit $\tau\to0$ these solution have the asymptotic form
(\ref{lim0}) for $k=\pm1$ or (\ref{liminf}) for $k=-1$.

We'll see below that the behavior of the mentioned solutions (\ref{asqrt})
and (\ref{asqn}) in the $t\to\infty$ limit radically differs from that
of all other solutions in the positive curvature case $k=1$, $\L>0$.
Otherwise, in the case $k=-1$, $\L>0$ all solutions have the asymptotic form
(\ref{asqn}).

Analyzing the solutions of the system (\ref{aqs}) we are to note that there
is the domain
\be
nq^2+\L a^2-kn(n-1)<0,
\label{neg}\ee
in the phase $aq$-plane where the radicand in (\ref{aqs}) or (\ref{qgen}) is
negative. Hence, the graphs of the solutions in phase plane can not pass
through the domain (\ref{neg}) and cross its border --- the 2-nd order curve.

For $k=1$ and an arbitrary fixed value $\L$ the graph of an every solution
of Eqs.~(\ref{aqs}) in the phase $aq$-plane contacts with the mentioned
border of the domain (\ref{neg}), that is this border (the ellipse if $\L>0$
or the hyperbola if $\L<0$) is the envelope of the graphs family. At the
contact point (we denote its coordinates $a_0$, $q_0$) for each solution
the sign before the radical in Eq.~(\ref{aqs}) changes from ``$-$" to ``$+$".
So any graph of solution in the case $k=1$ consists of two parts corresponding
to these signs and smoothly connected at the point ($a_0,q_0$).

In Fig.~3 the graphs of the considered solutions with $k=1$ in the $ta$-plane
are represented for the case $n=3$ and various $\L$ (Fig.~3a), and for the
case  $n=4$ with fixed value $\L=1$ (Fig.~3b). The parameter $t_0$ in
Eq.~(\ref{tau}) is chosen so the value $a=0$ corresponds to $t=0$. Under these
conditions the curves in Fig.~3b are the one parameter family, and one may
choose the value $a_0$ (if $q_0\ge0$) as the parameter numerating the curves
in the family. The values $a_0$ are shown near corresponding lines in Fig.~3b.
The curves marked by symbols ``4"  and ``5" in Fig.~3b correspond to the
values $q_0<0$ at the point of sign reversal in Eq.~(\ref{aqs}); these numbers
are maximal values of the scale factor $a$ for these solutions.

In Fig.~3a for the case $n=3$ the solutions $a=a(t)$ with the same value
$a_0=1$  for various values $\L$ (marked near the graphs) are shown.
For the curve with $\L=10$ the value $a=1$ is the maximum of $a$.
We see that presence of $\L<0$ accelerates the expansion in comparison with
the case $\L=0$ (Fig.~1) and $\L>0$ --- decelerates. But with increasing $\L$
(without exceeding the certain limit) the expansion remains irreversible ---
the values $a$ and $\phi$ grows up to $+\infty$ during the finite time
$T(a_0,\L)$. The asymptotic behavior $a(t)$ at $t\to T$ takes the form
(\ref{liminf}) or
\be
a\simeq{\mbox{const}}\cdot(T-t)^{-1/\sn},\quad
e^{2\phi}\simeq{\mbox{const}}\cdot(T-t)^{-1-\sn},\quad t\to T.
\label{liminfT}\ee
If $\L$ reaches some critical value (depending on $a_0$) the expansion
becomes unlimited in time with the power-law behavior (\ref{asqn}) at
$t\to\infty$. In Fig.~3a ($n=3$, $a_0=1$) this critical value slightly
exceeds $\L=2.732$. For larger $\L$ the type of evolution changes ---
the expansion in some time interval is replaced by compression that
is finished in the finite time $T=T(a_0,\L)$ with vanishing $a$.
The value $\phi(t)$ during the time $T$ grows from $-\infty$ up to some
maximum and then decreases to $-\infty$. In the limit $t\to T$ the asymptotic
relation (\ref{lim0}) takes place.

Fig.~3b illustrates that the critical power-law solution (it takes the form
(\ref{asqrt}) for $n=4$) may be obtained for fixed value $\L>0$ by choosing
the value $a_0$. Any small deviation from the critical value results in
an evolution within the framework of any of the two mentioned types with finite lifetime $T$.

For the case of negative spatial curvature $k=-1$ and $\L<0$ each graph
of solution in the phase $aq$-plane contacts with the border of the
forbidden domain (\ref{neg}) with the sign reversal in Eq.~(\ref{aqs})
(similarly to the case $k=1$). The family of these graphs $a=a(t)$ for
fixed $n=4$, $\L=-1$ and various values $a_0$ is shown in Fig.~4a.
The coordinates of the contact point $a_0$ numerate the curves as in Fig.~3b.

All the solutions with $\L<0$ (Fig.~4a) have finite lifetimes $T(a_0,\L)$
and the behavior (\ref{liminfT}). But their behavior at $t\to0$ depends on
$a_0$ and $\L$. If $a_0$ does not exceed some critical value $a_0=a_{cr}(\L)$ then
$a,\phi\to+\infty$ at $t\to0$ in accordance with (\ref{liminf}) (similarly
to the case $\L=0$ in Fig.~2b). In the opposite case $a_0>a_{cr}$ this limit is
$a,e^{2\phi}\to 0$ (\ref{lim0}) (compare with Fig.~2a). These two cases are
divided by the solution with the critical value $a_0=a_{cr}$ (Fig.~4a) that is
close to the linear expansion (\ref{alin}) $a\simeq t$, $\phi\simeq\phi_0$ for
$t\ll T$.

In the case $k=-1$, $\L>0$ there is no real forbidden domain (\ref{neg})
but the solutions with different signs in Eq.~(\ref{aqs}) are connected by the
time reversal $t\to-t$, $q\to-q$. So we have one family of solutions with
the infinite lifetime and $a\sim\sqrt\tau$ (\ref{aqs}) asymptotic behavior
at $t\to\infty$ \cite{Tseyt92}, that is shown in Fig.~4b for $\L=1$ and $n=0$.
The values $q=\dot a$ at the point $a=a_0=1$ numerate the curves. For values
$q<q_{cr}(a_0,\L)$ (here $q_{cr}\simeq1$) the solutions have the $t\to0$ limit
(\ref{liminf}) and for $q>q_{cr}$ --- the limit (\ref{lim0}). The critical
solution with $q=q_{cr}$ (Fig.~4b) has $a\simeq t$, $\phi\simeq\phi_0$ dependence for small $t$.

\bigskip
\noindent{\large\bf Conclusion}
\medskip

The Friedmann type cosmological solutions in $D$\,-\,dimensional
dilaton gravity with the action (\ref{S}) are investigated in this paper.
Different values of spatial curvature $k=0,\,\pm1$ and various values $\L$
are considered. For all cases the solutions of the system
(\ref{eq1})\,--\,(\ref{eq3}) (cosmological solutions) are described and
classified. They include as the expressions (\ref{ak0})\,--\,(\ref{phik0}), (\ref{lin}), (\ref{alin}) (particular cases or generalizations of them were
obtained in Refs.~\cite{BailinL}\,--\,\cite{Tseyt92}), as new analytic solutions
(\ref{dt})\,--\,(\ref{phip}), (\ref{n4}), (\ref{asqrt}).

To classify various types of solutions we represent them in the following table:
\begin{center}
Table 1. Types of cosmological solutions.\\
\begin{tabular}{|c|c|c|c|} \hline
 &$\L<0$ & $\L=0$ & $\L>0$\\ \hline
 &  & & $0\;\to\;\infty$\\
$k=1$ & $0\;\to\;\infty$ & $0\;\to\;\infty$ & $0\;\to\,a\simeq a_0\sqrt\tau$\\
 &  & & $0\;\to\;0$\\ \hline
 &  &$\infty\;\to\,a\simeq a_0\tau^{-1/\sn}$ & $\infty\;\to\,a\simeq a_0$\\
$k=0$ & $0\;\to\;\infty$ & $a=a_0,\;\phi=\phi_0$ & $a=a_0,\;\phi=
\frac12\sqrt\L\,\tau$\\
 &  & $0\;\to\,a\simeq a_0\tau^{1/\sn}$ & $0\;\to\,a\simeq a_0$\\ \hline
& $\infty\;\to\;\infty$ & $\infty\;\to\,a\simeq\tau$ & $\infty\;\to
\,a\simeq a_0\sqrt\tau$\\
$k=-1$ & $a\simeq\tau\;\to\;\infty$ & $a=\tau,\;\phi=\phi_0$ & $a\simeq\tau\;
\to\,a\simeq a_0\sqrt\tau$\\
& $0\;\to\;\infty$ & $0\;\to\,a\simeq\tau$ & $0\;\to\,a\simeq a_0\sqrt\tau$\\
\hline
\end{tabular}
\end{center}
Here the symbol ``$0$"  denotes the beginning or the end\footnote{Remind that
we consider two solutions connected by the time reversal as identical ones.}
of an evolution with the asymptotic behavior (\ref{lim0}) ($a,e^{2\phi}\to 0$).
The symbol ``$\infty$" corresponds to the asymptotics $a,\phi\to+\infty$ of the
type (\ref{liminf}) or (\ref{liminfT}). Note that for $\L<0$ a lifetime of
the dilatonic universe is always finite --- the beginning and the end of the
evolution are marked in Table~1 by the mentioned symbols
(for $k=-1$ we have the critical solution with close to linear $a\simeq\tau$, $\phi\simeq\phi_0$ beginning of the expansion and the finite lifetime).

Other symbols describing the end of an evolution in Table~1 correspond to
an infinite lifetime and power-law evolution within one of the following
types: (\ref{asqn}) ($a\sim\sqrt\tau$), (\ref{lin}),
(\ref{ak0})\,--\,(\ref{phik0}) ($k=0$, $\L\ge0$), (\ref{alin}),
(\ref{dt})\,--\,(\ref{phip}) ($k=-1$, $a\sim\tau$).

For $k=-1$, $\L>0$ all solutions have the asymptotics $a_0\sqrt\tau$
(\ref{asqn}) \cite{Tseyt92}, but for $k=1$, fixed $\L>0$ and $t_0$
only one critical solution (dividing different types of solutions)
has such a behavior.

Note that the considered cosmological solutions have identical qualitative
features for various dimensionalities $n\ge2$ ($n=1$ or $D=2$ is reduced to the
case $k=0$). In this sense the classification in Table~1 is universal for all $D$.

\end{document}